\documentclass[12pt]{article}
\usepackage{amsmath, amsthm, amssymb,amsfonts}

\usepackage{hyperref}
\begin{document}
\begin{titlepage}
\begin{center}
{\bf\Large{\vbox{\centerline{Entanglement temperature for black branes} \vskip .5cm
 \centerline{with hyperscaling violation}}}}
\vskip
0.5cm
{Xiao-Bao Xu$^{a}$ \footnote{xbxu789@163.com}, Gu-Qiang Li$^{a}$ \footnote{zsgqli@hotmail.com}, Jie-Xiong Mo$^{a}$ \footnote{mojiexiong@gmail.com}} \vskip 0.05in
{\it ${}^a$ Institute of Theoretical Physics,\\ Lingnan Normal University, Zhanjiang, 524048 Guangdong, China }
\end{center}

\vskip 0.5in
\baselineskip 16pt
\abstract{Entanglement temperature is an interesting quantity which relates the increased amount of entanglement entropy and energy for a weakly excited state in entanglement first-law, it is proportional to the inverse of the size of the entanglement subsystem and only depends on the shape of the entanglement region. We find the explicit formula of entanglement temperature for the general hyperscaling violation backgrounds with a strip-subsystem.  We then investigate the entanglement temperature for a round ball-subsystem, we check that the entanglement temperature has a universal form when the hyperscaling violation exponent is near zero.}
\end{titlepage}

\section{Introduction}
The process of thermalization of a nonequilibrium state is a common phenomenon in the nature, in recent years we can study such process in laboratory experiments. However we still lack efficient tools to make insights into the behavior of far-from-equilibrium system. One of difficulties in the process of thermalization is that the thermodynamical quantities such as temperature, entropy, pressure, etc., may not be well defined when the system is out of equilibrium. Nevertheless, there are still useful quantities to probe the nonequilibrium system, for example entanglement entropy. But due to strongly coupled interactions during the process of thermalization, we can't compute the entanglement entropy easily in the quantum field theory. It is remarkable that gauge/gravity duality \cite{Maldacena97} provides us a general theoretical tool to explore non-equilibrium properties of strongly coupled field theories, see for instance \cite{Chesler0809}.

According to AdS/CFT correspondence \cite{Maldacena97,Gubser98,Witten98}, Ryu and Takayanagi \cite{rt} find that the entanglement entropy of a particular spatial region V in the boundary theory could be given by the formula $S=\frac{\mathcal A}{4G_N}$ in the bulk, where $\mathcal A$ is the area of the minimal 
surface whose boundary is given by $\partial V$ . Their proposal has been checked and extended in many ways, this relation is proven in \cite{Casini} for a spherical entangling region and for more general case in \cite{Lewkowycz13}, quantum corrections to this area law formula is considered in \cite{Barrella,Faulkner} and holographic entanglement entropy in high derivative gravity theory has been obtained in \cite{Hung,deBoer,Dong,Camps,Miao}.

Recently, in \cite{Bhattacharya} the authors show that for a sufficiently small subsystem, the variation of the entanglement entropy is proportional to the variation of the energy of the subsystem for an excited state around the vacuum state. In particular, the proportionality constant is related to the size of the entanglement region. We then get a first law in analogy with the first law of thermodynamics by identifying the proportionality constant as the inverse of the so-called entanglement temperature, this relation is viewed as the first law of entanglement thermodynamics \cite{Allahbakhshi}. Subsequently the entanglement temperature in high derivative gravity is investigated in \cite{Guo} and entanglement thermodynamics in the Lifshitz geometry is analyzed in \cite{Chakraborty}.

In this manuscript we would like to find the explicit expression of entanglement temperature in the hyperscaling violating geometry, which is still a gap in the literature. The metric of the hyperscaling violating geometry is as follows
\begin{equation}
ds^2=r^{\frac{2\theta}{d}} \left(-\frac{1}{r^{2z}}dt^2+\frac{1}{r^2}dr^2+\frac{1}{r^2}d{\bf x}^2 \right), \label{metric}
\end{equation}
where $z$ and $\theta$ are dynamical and hyperscaling violation exponents respectively. This metric is not scale invariant under the scale transformation
\begin{equation}
t\rightarrow \lambda^z t,\;\;\;\;\;\;\;r\rightarrow \lambda r,\;\;\;\;\;\;\;{\bf x}\rightarrow \lambda {\bf x},\;\;\;\;\;\;\;
ds \rightarrow \lambda^{\frac{\theta}{d}} ds.
\end{equation}
The gravity solution of this kind have been found in Einstein-Maxwell-Dilaton(EMD) theory \cite{emd,emd1,emd2,emd3}, some properties of the hyperscaling violating geometry have been studied in \cite{Dong,Hartnoll12,Bai,Xu,Alishahiha14,Gouteraux14,Fonda14}.

One of interesting features of the above metric is that for the case $\theta=d-1$, the holographic entanglement entropy exhibits a logarithmic
violation of area law \cite{Ogawa,Huijse}, indicating that the metric \eqref{metric1} could be a gravitational background dual to a theory with an $\mathcal O(N^2)$ Fermi surface, where $N$ is the number of degrees of freedom.

In this paper we first construct a simple black brane with hyperscaling violation by reducing the result of \cite{emd3}, which is considered as the gravity dual to an excited state around the zero temperature system, i.e. the ground state with the corresponding metric \eqref{metric1}. The hyperscaling violating geometry are not asymptotically $AdS$ spacetimes, so there are no well defined Fefferman-Graham asymptotic expansions and no covariant local boundary counter-terms to the bulk action \cite{Henningson,Balasubramanian}. Holographic renormalization of non-relativistic backgrounds has been studied in many papers \cite{Taylor,Ross,Ross1,Baggio,Mann,Chemissany}, based on these works, we can get the holographic stress tensor for the hyperscaling violating geometry in the EMD theory \cite{Dehghani}. Then we can get explicitly the entanglement temperature for the strip-subsystem and also show that the entanglement temperature of a round ball-subsystem obeys the universal form $T_{ent}\propto \textit{l}^{-z}$ which has been argued in \cite{Bhattacharya}.

The paper is organized as follows. In the next section we obtain a simple black brane with hyperscaling violation in an EMD theory, and we calculate the holographic stress tensor for this model. In section three we make some computations on the entanglement temperature. Then we conclude in the last section.

\section{EMD theory and holographic stress tensor}
In this section we will briefly derive the black brane with hyperscaling violation following the procedure of \cite{emd3} and the associated holographic stress tensor of this model is analyzed by generalizing the method of \cite{Ross} for the case of Lifshitz spacetimes  to the hyperscaling violating geometry, indeed this has been done in \cite{Dehghani} partially. We will find the result of \cite{Dehghani} is sufficient for our aim.

To get a black brane with hyperscaling violation and no charge freedom, we consider the following minimal model
\begin{equation}
\label{action}
S=\frac{1}{16\pi G}\int d^{d+2}x\sqrt{-g}\left[R-\frac{1}{2}(\partial\phi)^2+V(\phi)-\frac{1}{4}
e^{\lambda \phi}F^2\right].
\end{equation}
The potential is taken as follows
\begin{equation}
V=V_0e^{\gamma\phi},
\end{equation}
here $\lambda, \gamma$ and $V_0$ are free parameters of the model.   \\
The equations of motion of the above action can be written down directly from \cite{emd3}. And we still consider the ansatz for the metric, scalar and gauge field in \cite{emd3} for our case,
\begin{equation}
ds^2=r^{2\alpha}\left(-r^{2z}f(r)dt^2+\frac{dr^2}{r^2f(r)}+r^2d\vec{x}^2\right),\;\;\;\;\phi=\phi(r),
\;\;\;\;\; F_{rt}\neq 0,
\end{equation}
and assume that the other components of gauge fields are  zero.
Then we solve the equations of motion by the approach of \cite{emd3}, there are two interesting geometries with hyperscaling violation, one is vacuum solution, the other is black brane type. More precisely, the metric of hyperscaling violating spacetime
\begin{equation}
ds^2=r^{2\alpha}\left(-r^{2z}dt^2+\frac{dr^2}{r^2}+r^2d\vec{x}^2\right). \label{metric1}
\end{equation}
In order to get a physically sensible dual field theory, we should impose the null energy conditions (NEC) at least, which lead to the following constraints on the solutions of Einstein's equations \cite{Dong,emd3}
\begin{align}
(\alpha+1)(\alpha+z-1)\geq 0,\cr
(z-1)(d(1+\alpha)+z)\geq 0.
\end{align}
For the simplicity of the discussion, we also assume that $\alpha+z\geq 0$ and $\alpha+1 \geq 0$, then the boundary of the metric \eqref{metric1}, i.e. $r\rightarrow \infty$, describes the UV of the dual field theory \cite{Dong}. The matter parts are as follows
\begin{align}
F_{rt}&=e^{-\lambda \phi}\;r^{\alpha(2-d)+z-d-1}\rho\cr
e^{\phi}&=e^{\phi_0}r^{\sqrt{2d(\alpha+1)(\alpha+z-1)}},
\end{align}
and the parameters $\lambda, \rho$ are fixed by
\begin{equation}
\lambda= -\frac{2\alpha(d-1)+2d}{\sqrt{2d(\alpha+1)(\alpha+z-1)}},\;\;\;
\rho^2=\frac{2V_0(z-1)e^{-\sqrt{\frac{2d(\alpha+1)}{\alpha+z-1}}\phi_0}}{d\,\alpha+d+z-1}.
\end{equation}
To get the hyperscaling violation solution, we should also require that
\begin{equation}
\gamma=\frac{-2\alpha}{\beta},\;\;\;\;{V_0}=e^{\frac{2\alpha{\phi_0}}{\sqrt{2d(1+\alpha)(-1+z+\alpha)}}}
(d\, \alpha+z+d-1 ) (d\,\alpha+z+d  ).
\end{equation}
Here we denote that $\beta=\sqrt{2d(\alpha+1)(\alpha+z-1)}$. Through a further reduction, the  hyperscaling violating spacetime is given as follows
\begin{align}\label{matter}
ds^2&=r^{2\alpha}\left(-r^{2z}dt^2+\frac{dr^2}{r^2}+r^2d\vec{x}^2\right),\cr
F_{rt}&=\sqrt{2(z-1)(z+d+d\,\alpha)}e^{\frac{\alpha(d-1)+d}{\beta}\phi_0}\;r^{d+z+\alpha\, d-1},\cr
A_{t}&=\sqrt{\frac{2(z-1)}{z+d+d\alpha}}e^{\frac{\alpha(d-1)+d}{\beta}\phi_0}\;r^{d+z+\alpha\, d},\cr
e^{\phi}&=e^{\phi_0}r^{\sqrt{2d(\alpha+1)(\alpha+z-1)}}.
\end{align}
The action \eqref{action} also admits a black brane with hyperscaling violation as a solution \cite{emd3}
\begin{equation}
ds^2=r^{2\alpha}\left(-r^{2z}f(r)dt^2+\frac{dr^2}{r^2f(r)}+r^2d\vec{x}^2\right), \label{hvblack}
\end{equation}
where
\begin{equation}\label{blackone}
f(r)=1-\frac{m}{r^{z+d+d\,\alpha}}.
\end{equation}
The corresponding Hawking temperature is found
\begin{equation}
T=\frac{\hat{\kappa}}{2\pi}=\frac{(z+d+d\,\alpha)r_H^z}{4\pi},
\end{equation}
where $r_H$ is the radius of horizon.

The gauge field and dilaton are almost the same as in the the zero temperature solution
\begin{align}\label{matter1}
A_{t}&=\sqrt{\frac{2(z-1)}{z+d+d\,\alpha}}e^{\frac{\alpha(d-1)+d}{\beta}\phi_0}\;(r^{d+z+\alpha\, d}-r_H^{d+z+\alpha\, d}),\cr
e^{\phi}&=e^{\phi_0}r^{\sqrt{2d(\alpha+1)(\alpha+z-1)}}.
\end{align}
We fixed the integration constant such that the gauge field vanishes on the horizon, leading to a regular gauge field.

Now we want to calculate the holographic stress tensor of the geometry \eqref{hvblack}. To do this, we need adding some counter-terms to the action \eqref{action}. The author of \cite{Ross} introduced some counter-terms to get the finite on-shell action for the Lifshitz solution. We will generalize this result to the hyperscaling violating geometry. We would like to consider the following counter-terms to make the action finite \cite{Dehghani}
\begin{equation}
S_{ct}=-\frac{1}{16\pi G}\int d^{d+1}\xi\sqrt{-h}r^{-\alpha}(c_0+c_1(-e^{\lambda \phi}A_\gamma A^\gamma)^{1/2}).
\end{equation}
The total action is given by adding the Gibbons-Hawking boundary term and the counter-term
\begin{eqnarray}
S_{tot}&=&\frac{1}{16\pi G}\int d^{d+2}x\sqrt{-g}\left[R-\frac{1}{2}\left(\partial\phi\right)^2+V\left(\phi\right)-\frac{1}{4}e^{\lambda\phi}F^2\right]+
\frac{1}{8\pi G}\int d^{d+1}\xi\sqrt{-h}K\nonumber\\&&-\frac{1}{16\pi G}\int d^{d+1}\xi\sqrt{-h}r^{-\alpha}(c_0+c_1(-e^{\lambda \phi}A_\gamma A^\gamma)^{1/2}),
\end{eqnarray}
where $K_{ab}=\nabla_a n_b$ is the extrinsic curvature of the boundary, the unit vector $n_a$ is orthogonal to the boundary.

Therefore the variation of the total action about an on-shell solution is just a boundary term
\begin{eqnarray}
\delta S_{tot}&=&\frac{1}{16\pi G}\int d^{d+1}\xi \sqrt{-h}\left[ \qquad  \right.\nonumber\\&&\left.
       \left(\pi_{ab}+\frac{1}{2}h_{ab}r^{-\alpha}\left(c_0+c_1\left(-e^{\lambda \phi}A_\gamma A^\gamma\right)^{1/2}\right)
           -c_1 r^{-\alpha}\frac{1}{2}\left(-e^{\lambda \phi}A_\gamma A^\gamma\right)^{-1/2}\left(-e^{\lambda \phi}A_{a} A_{b}\right)
\right)\delta h^{ab} \right.\nonumber\\&& \left.
-\left(n_a e^{\lambda \phi}F^{ab}+c_1 r^{-\alpha}\frac{1}{2}\left(-e^{\lambda \phi}A_\gamma A^\gamma\right)^{-1/2}\left(-e^{\lambda \phi}2A^b\right)\right)\delta A_b \right.\nonumber\\&& \left.
-\left(n_a\nabla^a\phi+c_1 r^{-\alpha}\frac{1}{2}\left(-e^{\lambda \phi}A_\gamma A^\gamma\right)^{-1/2}\left(-\lambda e^{\lambda \phi}A_\gamma A^\gamma\right)\right)\delta \phi\right ],
\end{eqnarray}
where $\pi_{ab}=K_{ab}-K h_{ab}$.

To our interest in holographic renormalization of the hyperscaling violating geomerty, we just need focusing on the coefficients of $\delta h^{ab}$ and $\delta A_a$, it is the reason that $\delta \phi$ is always vanish in our case, this can be seen in \eqref{matter} and \eqref{matter1}.

At first we should make sure that the action satisfies $\delta S_{tot}=0$ for arbitrary variation around \eqref{metric1}, this allows us to determine the parameters $c_0, c_1$ in the counter-terms,
\begin{equation}
c_0=2d(1+\alpha),\;\;\;\;\;\;\;c_1=\sqrt{2(z-1)(z+d+d\,\alpha)}.
\end{equation}
We could also find $S_{tot}=0$ for the background \eqref{metric1}.

Then the general variation of the action can be written as
\begin{equation}
\delta S_{tot}=\int d^{d+1}\xi\left\{s_{ab}\delta h^{ab}+s_a\delta A^a\right\},
\end{equation}
where
\begin{eqnarray}
s_{ab}&=&\frac{\sqrt{-h}}{16\pi G} \left[\pi_{ab}+r^{-\alpha}h_{ab}d \left(1+\alpha\right)+\frac{1}{2}\sqrt{2\left(z-1\right)\left(z+d+d\,\alpha\right)}\,r^{-\alpha}e^{\frac{\lambda\phi}{2}}\left(-A_\gamma A^\gamma\right)^{-1/2}\right.\nonumber\\&& \left.\left(A_a A_b-A_\gamma A^\gamma h_{ab}\right)\right],\nonumber\\
s_b&=&-\frac{\sqrt{-h}}{16\pi G}\left[n^a e^{\lambda\phi}F_{ab}-\sqrt{2\left(z-1\right)(z+d+d\,\alpha)}\,r^{-\alpha}e^{\lambda\phi/2}(-A_\gamma A^\gamma)^{-1/2}A_b\right].
\end{eqnarray}

Here we will give the energy density and spatial stress tensor for the non-relativistic theory, which may be useful for getting the entanglement temperature. The expressions of the stress tensor are derived in \cite{Ross}
\begin{eqnarray}
&&{\mathcal E} = 2 s^t_{\ t} - s^t A_t, \quad
{\mathcal E}^i = 2s^i_{\ t} - s^i A_t,\nonumber\\
&&{\mathcal P}_i = -2 s^t_{\ i} + s^t A_i,  \quad \Pi_{i}^j =-2 s^j_{\
  i} + s^j A_i.
\end{eqnarray}
With the above formula, the energy density and one component of spatial stress tensor for our solution \eqref{hvblack} have the forms at leading order one
\begin{equation}\label{energy density}
{\mathcal E}=\frac{d(1+\alpha)m}{16\pi G},\;\;\;\;\;\;\Pi_x^x=\frac{m\,z}{16\pi G}.
\end{equation}
where $m$ is the blacken factor in \eqref{blackone}.

We find that the same formula of energy density has appeared in the paper \cite{Dehghani}, which studied the the thermodynamics of hyperscaling violating black branes in the presence of a nonlinear massless electromagnetic field. The spatial stress tensor has been identified as the entanglement pressure in the general entanglement first-law \cite{Allahbakhshi,Chakraborty}.

\section{Entanglement temperature}
In this section we will study holographic entanglement entropy of the excited state due to the metric perturbation \eqref{hvblack} and the corresponding  entanglement temperature in detail. Before doing the specific calculations, we convert the expression of our metric \eqref{hvblack}, such that the boundary of the new one is at $r\rightarrow 0$, i.e.
\begin{eqnarray}\label{inhvblack}
ds^2&=&r^{-2\alpha}\left(-r^{-2z}g(r)dt^2+\frac{dr^2}{r^2g(r)}+r^{-2}d\vec{x}^2\right),\nonumber\\
g(r)&=&1-m\,r^{z+d+d\,\alpha}.
\end{eqnarray}
\subsection{Entangling surface with a strip}
Let us consider an entangling region in the shape of a strip with the width of $\ell$ given by
\begin{equation}\label{strip}
-\frac{\ell}{2}\leq
x_1\leq \frac{\ell}{2},\;\;\;\;\;\;\; 0\leq x_i\leq
L,\;\;\;\;\;i=2,\cdots, d.
\end{equation}
We can parameterize the minimal surface $\gamma_A$ by $x_1=x(r)$, then its area is as follows
\begin{equation}
\mathcal A=2L^{d-1}\int^{r_*}_0 dr\, r^{-(\alpha+1)d}\sqrt{1+m\,r^{z+d+d\,\alpha}+x'^2},
\end{equation}
where $'$ denotes the derivative to $r$.

By making use of the standard procedure one may minimize the area to get \cite{rt}
\begin{eqnarray}
\mathcal A&=&2L^{d-1}\int^{r_*}_0 dr\, r^{-(\alpha+1)d}\sqrt{\frac{1+m\,r^{z+d+d\,\alpha}}{1-(\frac{r}{r_*})^{2(\alpha+1)d}}}\label{area}\\
\ell&=&2\int^{r_*}_0 dr\,(\frac{r}{r_*})^{(\alpha+1)d}\sqrt{\frac{1+m\,r^{z+d+d\,\alpha}}{1-(\frac{r}{r_*})^{2(\alpha+1)d}}}
\end{eqnarray}
where $r_*$ is the turning point of $\gamma_A$.

In the limit $m\,\ell^{z+d+d\,\alpha}\ll1$, we can expand \eqref{area} up to the first order of $m\,\ell^{z+d+d\,\alpha}$ and find that
\begin{equation}
\Delta \mathcal A=\mathcal A-\mathcal A^{(0)}=mL^{d-1}r_*^{z+1}\frac{\sqrt{\pi}\Gamma(\frac{1+z}{2(\alpha+1)d})}{2(1+(\alpha+1)d+z)\,\Gamma(\frac{1+(\alpha+1)d+z}{2(\alpha+1)d})},
\end{equation}
where $\mathcal A^{(0)}$ is the  area of minimal surface in the pure hyperscaling violating geometry \eqref{metric1}, its value is given in \cite{Dong}.

Here we assume that the perturbation do not change the shape of the entanglement surface in the pure background \eqref{metric1}, the reason of which can be seen in \cite{Guo}. The length of the strip is related to the turning point by
\begin{eqnarray}
\ell&=&2\int^{r_*}_0 dr\,(\frac{r}{r_*})^{(\alpha+1)d}\sqrt{\frac{1}{1-(\frac{r}{r_*})^{2(\alpha+1)d}}}\nonumber\\
&=&2r_*\frac{\sqrt{\pi}\Gamma(\frac{1+(\alpha+1)d}{2(\alpha+1)d})}{\Gamma(\frac{1}{2(\alpha+1)d})}.
\end{eqnarray}

Therefore the variation of holographic entanglement entropy can be written in terms of the length of the strip $\ell$ as
\begin{equation} \label{entropy}
\Delta S_E=\frac{\Delta \mathcal A}{4G}=\frac{m\,L^{d-1}}{4G}(\frac{\ell}{2})^{z+1}\frac{\Gamma\left(\frac{1}{2(\alpha+1)d}\right)^{z+1}}{\sqrt{\pi}^z\Gamma\left(\frac{1+(\alpha+1)d}{2(\alpha+1)d}\right)^{z+1}}
\frac{\Gamma(\frac{1+z}{2(\alpha+1)d})}{2(1+(\alpha+1)d+z)\,\Gamma(\frac{1+(\alpha+1)d+z}{2(\alpha+1)d})}.
\end{equation}
On the other hand the variation of energy in the subsystem $A$ is as follows
\begin{equation}\label{energy}
\Delta E_A=\int dx^{d}\,\mathcal E=L^{d-1}\ell\frac{d(\alpha+1)m}{16\pi G}.
\end{equation}
Unlike the result of \cite{Chakraborty}, there are no contributions of the chemical potential from Maxwell field, because we find $s_t=0$ in our perturbation.
Then from \eqref{entropy} and \eqref{energy}, we can find the entanglement temperature for the strip-subsystem has the following form
\begin{eqnarray}\label{entT}
\frac{1}{T_{ent}}&=&\frac{\Delta S_E}{\Delta E_A}\nonumber\\
&=&\frac{\pi^{1-z/2}\left(\frac{\ell}{2}\right)^z\Gamma\left(\frac{1}{2(\alpha+1)d}\right)^{z+1}\Gamma(\frac{1+z}{2(\alpha+1)d})}{(\alpha+1)d\,(1+(\alpha+1)d+z)\,\Gamma\left(\frac{1+(\alpha+1)d}{2(\alpha+1)d}\right)^{z+1}
\Gamma(\frac{1+(\alpha+1)d+z}{2(\alpha+1)d})}.
\end{eqnarray}
Thus the explicit entanglement temperature of the hyperscaling violating geometry for the strip-subsystem is proportional to $\ell^{-z}$, which has been already anticipated in \cite{Bhattacharya}. However the authors of \cite{Bhattacharya} use the thermal entropy of the hyperscaling violating black brane to get the energy density, this is different from our approach. And our formula \eqref{entT} can reduce to the one in \cite{Bhattacharya} when our geometry is the AdS one.

\subsection{Entangling surface with a sphere}
Let us choose the subsystem to be a round ball with radius $R_0$. The minimal surface can be parameterized by $r(u)$ and $\rho(u)$, where $\rho$ is the radial direction of the ball. So the induced metric on the codimension two hypersurface in the bulk is given by
\begin{equation}
ds_D^2=r^{-2-2\alpha}\left[(1+m\,r^{z+d+d\,\alpha})\dot{r}^2du^2+\dot{\rho}^2du^2+\rho^2d\Omega^2_{d-1}\right],
\end{equation}
where the dot denotes the derivative with respect to $u$. So the area of the hypersurface in the bulk is
\begin{equation}
\mathcal A=\mathrm Vol(S^{d-1})\int du\,r^{-(\alpha+1)d}\rho^{d-1}\sqrt{(1+m\,r^{z+d+d\,\alpha})\dot{r}^2+\dot{\rho}^2}.
\end{equation}
Here we adopt the following ansatz to find the minimal surface \cite{Hung,Guo}
\begin{equation}\label{coordtrs}
\rho(u)=f(\frac{u}{R_0})\mathrm{cos}(\frac{u}{R_0}),\;\;\;\;\;r(u)=f(\frac{u}{R_0})\mathrm{sin}(\frac{u}{R_0}),\;\;\;\varepsilon\leq u \leq \frac{\pi}{2}R_0.
\end{equation}
The holographic entanglement for pure hyperscaling violating geometry is
\begin{equation}\label{action1}
\mathcal A^{(0)}=\frac{Vol(S^{d-1})}{4G}\int^{\pi/2}_{\varepsilon/R_0}dx\left(\frac{\mathrm{cos}(x)}{\mathrm{sin}^{\alpha+1}(x)}\right)^{d-1}\frac{1}{\mathrm{sin}^{\alpha+1}(x)}f^{-\alpha\,d}(x)\sqrt{1+(\frac{d\,\mathrm{ln}f}{dx})^2}.
\end{equation}
The equation of motion of the holographic entanglement entropy \eqref{action1} is equivalent to the following one
\begin{multline}\label{eom}
(\alpha+1)d\csc(x)f'(x)\left(f'(x)^2+f(x)^2\right)+(d-1)\tan(x)\sec(x)f'(x)\left(f'(x)^2+f(x)^2\right) \\-f(x)\sec(x)\left((\alpha\,d-1) f'(x)^2+\alpha\,d \,f(x)^2+f(x) f''(x)\right)=0.
\end{multline}
Because of the hyperscaling violation exponent $\alpha$, $f(x)=const.$ is not a solution of the equation \eqref{eom}. However if we assume that $\alpha$ is close to 0, we can get the following solution up to the first order of $\alpha$,
\begin{equation}
f(x)=c_0+\alpha \,g(x),
\end{equation}
where we will set $c_0=R_0$, and $g(x)$ is an analytic function, which is obtained by Mathematica
\begin{eqnarray}
g(x)&=& C_1+\frac{C_2 \cos ^{2-d}(x) \, _2F_1\left[\frac{1}{2}-\frac{d}{2},1-\frac{d}{2};2-\frac{d}{2};\cos ^2(x)\right]}{2-d}-\frac{1}{2(d-2)}c_0 \cos^2(x)\nonumber\\
&&\left(-2 \, _2F_1\left(\frac{1}{2}-\frac{d}{2},1-\frac{d}{2};2-\frac{d}{2};\cos ^2(x)\right)\, _2F_1\left(\frac{d}{2},\frac{d+1}{2};\frac{d}{2}+1;\cos ^2(x)\right)+d\,\Gamma \left(2-\frac{d}{2}\right)\right.\nonumber\\&& \left.\text{HypergeometricPFQRegularized}\left[\left\{1,\frac{3}{2},1\right\},\left\{2-\frac{d}{2},2\right\},\cos^2(x)\right]\right),
\end{eqnarray}
where $C_1$ and $C_2$ are the integration constants, and we can say that $g(x)$ is also proportional to $c_0$, i.e. $f(x)\propto R_0$.

With the above observation, we can investigate the form of the variation of holographic entanglement entropy for the entangling surface with sphere
\begin{equation}
\Delta S_E=\frac{\Delta \mathcal A}{4G}=\frac{m\,\mathrm Vol(S^{d-1})}{8G}R_0\int^{\pi/2}_{\frac{\varepsilon}{R_0}}dx\frac{\rho^{d-1}r^z\dot{r}^2}{\sqrt{\dot{r}^2+\dot{\rho}^2}},
\end{equation}
with the dot denote the derivative to $u$ not $x$.

According to the formula \eqref{coordtrs}, $\dot{r},\,\dot{\rho}$ don't have the factor $R_0$, thus the variation of holographic entanglement entropy behaves as $\Delta S_E\propto m\,R_0^{d+z}$. On the other hand, for the variation of the energy we have $\Delta E_A\propto m\,R_0^d$. Thus the entanglement temperature $T_{ent}$ of the spherical entangling surfaces is proportional to $R_0^{-z}$, which is consistent with the result in \cite{Bhattacharya}.

\section{Conclusions and discussions}
In this manuscript we investigate the variation of holographic entanglement entropy for the black brane with hyperscaling violation, and find the explicit entanglement temperature for the strip entangling surface and make sure that the entanglement temperature for the sphere entangling surface obeys the universal form $T_{ent}\propto R_0^{-z}$. Because we just turn on the metric perturbation of $tt$ component, we got the simplest entanglement first-law. It would be interesting to consider the general perturbative hyperscaling violating black brane as the cases in the Lifshitz geometry \cite{Chakraborty} and in the non-conformal D-branes \cite{Pang}, we may find the more complete entanglement first-law. To realize it, we should solve the problem of holographic renormalization involving all fields in the EMD theory, the work of \cite{Chemissany} might be useful.

\section*{Acknowledgements}
X.B. Xu would like to thank the ``Holographic duality for condensed matter physics'' held at KITPC, BeiJing, China for their financial support during the conference.

\end{document}